# Spectral Estimation of Conditional Random Graph Models for Large-Scale Network Data

**Antonino Freno**    **Mikaela Keller***    **Gemma C. Garriga**    **Marc Tommasi**\*

INRIA Lille – Nord Europe, Villeneuve d'Ascq (France)
{antonino.freno, mikaela.keller, gemma.garriga, marc.tommasi}@inria.fr

## Abstract

Generative models for graphs have been typically committed to strong prior assumptions concerning the form of the modeled distributions. Moreover, the vast majority of currently available models are either only suitable for characterizing some particular network properties (such as degree distribution or clustering coefficient), or they are aimed at estimating joint probability distributions, which is often intractable in large-scale networks. In this paper, we first propose a novel network statistic, based on the Laplacian spectrum of graphs, which allows to dispense with any parametric assumption concerning the modeled network properties. Second, we use the defined statistic to develop the *Fiedler random graph* model, switching the focus from the estimation of joint probability distributions to a more tractable conditional estimation setting. After analyzing the dependence structure characterizing Fiedler random graphs, we evaluate them experimentally in edge prediction over several real-world networks, showing that they allow to reach a much higher prediction accuracy than various alternative statistical models.

## 1 INTRODUCTION

Arising from domains as diverse as bioinformatics and web mining, large-scale data exhibiting network structure are becoming increasingly available. Network models are commonly used to represent the relations among data units and their structural interactions. Recent studies, especially targeted at social network modeling, have focused on random graph models of those networks. In the simplest form, a social network is a configuration of binary random variables $X_{uv}$ such that the value of $X_{uv}$ stands for the presence or absence of a link between nodes $u$ and $v$ in the network. The general idea underlying random graph modeling is that network configurations are generated by a stochastic process governed by specific probability laws, so that different models correspond to different families of distributions over graphs.

The simplest random graph model is the Erdős-Rényi (ER) model [Erdős and Rényi, 1959], which assumes that the probability of observing a link between two nodes in a given graph is constant for any pair of nodes in that graph, and it is independent of which other edges are being observed. In preferential attachment models [Barabási and Albert, 1999], the probability of linking to any specified node is proportional to the degree of the node in the graph, leading to "rich get richer" effects. Small-world models [Watts and Strogatz, 1998] try to capture instead some phenomena often observed in real networks such as high clustering coefficients and small diameters [Newman, 2010]. A sophisticated attempt to model complex dependences between edges in the form of Gibbs-Boltzmann distributions is made by exponential random graph (ERG) models [Snijders et al., 2006], which subsume the ER model as a special case. Finally, a recent attempt at modeling real networks through a stochastic generative process is made by Kronecker graphs [Leskovec et al., 2010], which try to capture phenomena such as heavy-tailed degree distributions and shrinking diameter properties while paying attention to the temporal dynamics of network growth.

While some of these models behave better than others in terms of computational tractability, one basic limitation affecting all of them is what we may call a *parametric assumption* concerning the probability laws underlying the observed network properties. In other words, currently available models of network structure assume that the probability distribution gener-ating the network can be expressed through a particular parametric form $P(\mathcal{G}|\boldsymbol{\theta})$, where $\mathcal{G}$ is the observed graph and $\boldsymbol{\theta}$ is a parameter vector. For example, typical formulations of exponential random graph models assume that the building blocks of real networks are given by such structures as $k$-stars and $k$-triangles, with different weights assigned to different structures, whereas Kronecker graphs assume that the number of edges and the number of nodes in a given network are related by a densification power law with a suitable densification parameter. In such frameworks, estimating the model from data reduces to fitting the model parameters, whereas the model structure remains fixed from the very beginning. The problem is that, in order for a parametric model to deliver an accurate estimate of the distribution at hand, its prior assumption concerning the form of the modeled distribution must be satisfied by the given data, which is something that we can rarely assess a priori. To date, the knowledge we have concerning real-world network phenomena does not allow to assume that any particular parametric assumption is really capturing in depth any network-generating law, although some observed properties may happen to be modeled fairly well.

---

*Université Charles de Gaulle – Lille 3, Villeneuve d'Ascq (France).

The aim of this paper is twofold. On the one hand, we take a first step toward *nonparametric* modeling of random networks by developing a novel network statistic, which we call the *Fiedler delta* statistic. The Fiedler delta function allows to model different graph properties at once in an extremely compact form. This statistic is based on the spectral analysis of the graph Laplacian. In particular, it is based on the smallest non-zero eigenvalue of the Laplacian matrix, which is known as Fiedler value [Fiedler, 1973, Mohar, 1991]. On the other hand, we systematically adopt a *conditional* approach to random graph modeling, i.e. we focus on the conditional distribution of edges given some neighboring portion of the network, while setting aside the problem of estimating joint distributions. The resulting conditional random graph model is what we call *Fiedler random graph* (FRG). Roughly speaking, to model the conditional distribution of an edge variable $X_{uv}$ with respect to its neighborhood, we compute the difference in Fiedler values for the vicinity subgraph including or excluding the edge $\{u,v\}$. The underlying intuition is that the variations encapsulated in the Fiedler delta for each particular edge will give a measure of the role of that edge in determining the algebraic connectivity of its neighborhood. As part of our contributions, we theoretically prove that FRGs capture edge dependencies at any distance within a given neighborhood, hence defining a fairly general class of conditional probability distributions over graphs.

Experiments on the estimation of (conditional) link distributions in large-scale networks show that FRGs are well suited for estimation problems on very large networks, especially small-world and scale-free networks. Our results reveal that the FRG model supersedes other approaches in terms of edge prediction accuracy, while allowing for efficient computation in the large-scale setting. In particular, it is known that the computation of the Fiedler delta scales polynomially in the size of the analyzed neighborhood [Bai et al., 2000]. And the experiments show that even for small sizes of neighborhoods (which allow for extremely fast computation), our model regularly outperforms the alternative ones—which we reformulate here in terms of conditional models so as to allow for transparent comparison.

The paper is organized as follows. Sec. 2 reviews some preliminary notions concerning the Laplacian spectrum of graphs. The FRG model is presented in Sec. 3, where we also show how to estimate FRGs from data, and we analyze the dependence structures involved in FRGs. In Sec. 4 we review (and to some extent develop) a few alternative conditional approaches to random graph estimation, starting from some well-known generative models. All considered models are then evaluated experimentally in Sec. 5, while Sec. 6 draws some conclusions and sketches a few directions for further work.

## 2 GRAPHS, LAPLACIANS, AND EIGENVALUES

Let $\mathcal{G} = (\mathcal{V}, \mathcal{E})$ be an undirected graph with $n$ nodes. In the following we assume that the graph is unweighted with adjacency matrix $\mathbf{A}$. For each (unordered) pair of nodes $\{u,v\}$ such that $u \neq v$, we take $X_{uv}$ to denote a binary random variable such that $X_{uv} = 1$ if $\{u,v\} \in \mathcal{E}$, and $X_{uv} = 0$ otherwise. Since the graph is undirected, $X_{uv} = X_{vu}$.

The degree $d_u$ of a node $u \in \mathcal{V}$ is defined as the number of connections of $u$ to other nodes, that is $d_u = |\{v : \{u,v\} \in \mathcal{E}\}|$. Accordingly, the degree matrix $\mathbf{D}$ of a graph $\mathcal{G}$ corresponds to the diagonal matrix with the vertex degrees $d_1, \ldots, d_n$ on the diagonal. The main tools exploited by the random graph model proposed here are the graph Laplacian matrices. Different graph Laplacians have been identified in the literature [von Luxburg, 2007]. In this paper, we use consistently the *unnormalized graph Laplacian*, defined as $\mathbf{L} = \mathbf{D} - \mathbf{A}$. The basic facts related to the unnormalized Laplacian matrix can be summarized as follows:

**Proposition 1** (Mohar [1991]). *The unnormalized graph Laplacian* $\mathbf{L}$ *of an undirected graph* $\mathcal{G}$ *with* $n$

nodes has the following properties: (i) **L** *is symmetric and positive semi-definite; (ii)* **L** *has n non-negative, real-valued eigenvalues* $0 = \lambda_1(\mathcal{G}) \leq \ldots \leq \lambda_n(\mathcal{G})$; *(iii) the multiplicity of the eigenvalue 0 of* **L** *equals the number of connected components in the graph, that is, $\lambda_2(\mathcal{G}) > 0$ if and only if $\mathcal{G}$ is connected.*

In the following, the (algebraic) multiplicity of an eigenvalue $\lambda$ will be denoted by $M(\lambda, \mathcal{G})$. If the graph has one single connected component, then $M(0, \mathcal{G}) = 1$, and the second smallest eigenvalue $\lambda_2(\mathcal{G}) > 0$ is called *Fiedler eigenvalue*. The Fiedler eigenvalue provides insight into several graph properties: when there is a nontrivial spectral gap, i.e. $\lambda_2(\mathcal{G})$ is clearly separated from 0, the graph has good expansion properties, stronger connectivity, and rapid convergence of random walks in the graph [Mohar, 1991]. For example, it is known that $\lambda_2(\mathcal{G}) \leq \mu(\mathcal{G})$, where $\mu(\mathcal{G})$ is the edge connectivity of the graph (i.e. the size of the smallest edge cut whose removal makes the graph disconnected). Notice that if the graph has more than one connected component, then $\lambda_2(\mathcal{G})$ will be also equal to zero, thus implying that the graph is not connected. Without loss of generality, we abuse the term Fiedler eigenvalue to denote the smallest eigenvalue different from zero, regardless of the number of connected components. In this paper, by Fiedler value we will mean the eigenvalue $\lambda_{k+1}(\mathcal{G})$, where $k = M(0, \mathcal{G})$.

For any pair of nodes $u$ and $v$ in a graph $\mathcal{G} = (\mathcal{V}, \mathcal{E})$, we define two corresponding graphs $\mathcal{G}^+$ and $\mathcal{G}^-$ in the following way: $\mathcal{G}^+ = (\mathcal{V}, \mathcal{E} \cup \{\{u,v\}\})$, and $\mathcal{G}^- = (\mathcal{V}, \mathcal{E} \setminus \{\{u,v\}\})$. Clearly, we have that either $\mathcal{G}^+ = \mathcal{G}$ or $\mathcal{G}^- = \mathcal{G}$. A basic property concerning the Laplacian eigenvalues of $\mathcal{G}^+$ and $\mathcal{G}^-$ is the following [Mohar, 1991, Anderson and Morley, 1985, Cvetković et al., 1979]:

**Lemma 1.** *If $\mathcal{G}^+$ and $\mathcal{G}^-$ are two graphs with n nodes, such that $\{u,v\} \subseteq \mathcal{V}$, $\mathcal{G}^+ = (\mathcal{V}, \mathcal{E} \cup \{\{u,v\}\})$, and $\mathcal{G}^- = (\mathcal{V}, \mathcal{E} \setminus \{\{u,v\}\})$, then we have that:*

1. $\sum_{i=1}^{n} \lambda_i(\mathcal{G}^+) - \lambda_i(\mathcal{G}^-) = 2$;

2. $\lambda_i(\mathcal{G}^+) \leq \lambda_i(\mathcal{G}^-)$ *for any $i$ such that $1 \leq i \leq n$.*

## 3 FIEDLER RANDOM GRAPHS

Fiedler random graphs are defined in Sec. 3.1, while in Secs. 3.2 and 3.3 we deal with the problems of estimating them from data and characterizing their statistical dependence structure respectively.

### 3.1 CONDITIONAL DISTRIBUTIONS

Given the notions reviewed above, we introduce the Fiedler delta function $\Delta\lambda_2$, which is defined as follows.

Let $k = M(0, \mathcal{G}^+)$. Then,

$$\Delta\lambda_2(u, v, \mathcal{G}) = \lambda_{k+1}(\mathcal{G}^+) - \lambda_{k+1}(\mathcal{G}^-) \qquad (1)$$

In other words, for any pair of nodes $u$ and $v$ in graph $\mathcal{G}$, the Fiedler delta value of the pair $\{u, v\}$ in $\mathcal{G}$ is the (absolute) variation in the Fiedler eigenvalue of the graph Laplacian that would result from removing edge $\{u, v\}$ from $\mathcal{G}^+$.

Lemma 1 immediately implies the following proposition:

**Proposition 2.** *For any graph $\mathcal{G} = (\mathcal{V}, \mathcal{E})$ and any pair of nodes $u$ and $v$ such that $\{u, v\} \in \mathcal{E}$, we have that $0 \leq \Delta\lambda_2(u, v, \mathcal{G}) \leq 2$.*

*Proof.* Let $\mathcal{G}^- = (\mathcal{V}, \mathcal{E} \setminus \{\{u,v\}\})$ and $k = M(0, \mathcal{G})$. The theorem follows straightforwardly from Lemma 1, given that $\Delta\lambda_2(u, v, \mathcal{G}) = \lambda_{k+1}(\mathcal{G}) - \lambda_{k+1}(\mathcal{G}^-)$. □

Using the Fiedler delta statistic, we can proceed to define a Fiedler random graph. Given $\mathcal{G} = (\mathcal{V}, \mathcal{E})$, let $\mathcal{X}(\mathcal{G})$ denote the set of random variables defined on $\mathcal{G}$, that is $\mathcal{X}(\mathcal{G}) = \{X_{uv} \colon u \neq v \wedge \{u,v\} \subseteq \mathcal{V}\}$. Then:

**Definition 1.** *Given a graph $\mathcal{G} = (\mathcal{V}, \mathcal{E})$, suppose we associate a subgraph $\mathcal{G}_{uv}$ of $\mathcal{G}$ to each variable $X_{uv} \in \mathcal{X}(\mathcal{G})$. We say that $\mathcal{G}$ is a* Fiedler random graph *if, for any $X_{uv} \in \mathcal{X}(\mathcal{G})$, we have that $P(X_{uv} | \mathcal{X}(\mathcal{G}_{uv}) \setminus \{X_{uv}\}) = P(X_{uv} | \Delta\lambda_2(u, v, \mathcal{G}_{uv}))$.*

Concerning Definition 1, we stress the importance of two points. First, the key intuition motivating FRGs is to treat the Fiedler delta function as a real-valued random variable defined over graph configurations, and to exploit this random variable as a compact representation of those configurations. Second, FRGs are *conditional* models, i.e. they only define conditional distributions of edges in a random graph given the possible configurations of other edges. The model definition does not specify whether and how the considered conditional distributions can be combined to obtain a consistent factorization of the joint distribution of $\mathcal{X}(\mathcal{G})$. In order to obtain that result, the model should take into account the global dependence structure of $\mathcal{X}(\mathcal{G})$, but this goal is not pursued by FRG modeling. Two interesting questions that can be raised on the nature of the FRG model are then the following. First, how does the Fiedler delta statistic behave with respect to the Markov dependence property [Besag, 1974, Frank and Strauss, 1986]? Second, how useful can be the information enclosed in the Fiedler delta statistic for learning and mining applications over large-scale networks? One basic result related to the first question is presented in Sec. 3.3, whereas Sec. 5 will address the second point.

## 3.2 MODEL ESTIMATION

The goal of estimating a FRG from data is to obtain an estimate of the conditional distribution $P(X_{uv}|\mathcal{X}(\mathcal{G}_{uv}) \setminus \{X_{uv}\})$, defined over subgraphs $\mathcal{G}_{uv}$ of $\mathcal{G}$. Starting from Definition 1, this quantity can be manipulated as follows:

$$P(X_{uv}|\mathcal{X}(\mathcal{G}_{uv}) \setminus \{X_{uv}\}) = P(X_{uv}|\Delta\lambda_2(u,v,\mathcal{G}_{uv}))$$
$$= \frac{p(\Delta\lambda_2(u,v,\mathcal{G}_{uv})|X_{uv})P(X_{uv})}{p(\Delta\lambda_2(u,v,\mathcal{G}_{uv}))}$$
$$= \frac{p(\Delta\lambda_2(u,v,\mathcal{G}_{uv})|X_{uv})P(X_{uv})}{\sum_{x_{uv}} p(\Delta\lambda_2(u,v,\mathcal{G}_{uv})|x_{uv})P(x_{uv})} \quad (2)$$

In Eq. 2, $P(X_{uv})$ is the prior probability of observing an edge between any two nodes $u$ and $v$, whereas $p(\Delta\lambda_2(u,v,\mathcal{G}_{uv})|X_{uv})$ is the conditional density of the Fiedler delta $\Delta\lambda_2(u,v,\mathcal{G}_{uv})$ given the value of $X_{uv}$. Suppose we have a network $\mathcal{G} = (\mathcal{V}, \mathcal{E})$, and that our training sample is a dataset $\mathcal{D}$, defined as $\mathcal{D} = \{(x_{u_1 v_1}, \mathcal{G}_{u_1 v_1}), \ldots, (x_{u_n v_n}, \mathcal{G}_{u_n v_n})\}$. In other words, the training data are $n$ observations of node pairs $\{u_i, v_i\}$ such that, for each one of them, we know both the value of $X_{u_i v_i}$ and the configuration of $\mathcal{G}_{u_i v_i}$ in $\mathcal{G}$. A simple estimate for $P(X_{uv})$ can then be obtained by computing the proportion of linked/unlinked node pairs $\{u_i, v_i\}$ in $\mathcal{D}$. Let $\mathcal{D}_{x_{uv}}$ denote the set $\{\{u_i, v_i\}: (x_{u_i v_i}, \mathcal{G}_{u_i v_i}) \in \mathcal{D} \land x_{u_i v_i} = x_{uv}\}$. Then, $\widehat{P}(x_{uv}) = \frac{1}{n}|\mathcal{D}_{x_{uv}}|$. On the other hand, in order to estimate $p(\Delta\lambda_2(u,v,\mathcal{G}_{uv})|X_{uv})$, we need to make a choice concerning the form of the modeled density function. In the lack of prior knowledge concerning the form of the density, a reasonable choice is to adopt a nonparametric estimation technique. In particular, we use a kernel density estimator [Rosenblatt, 1956, Parzen, 1962], defined as follows:

$$\hat{p}(\Delta\lambda_2(u,v,\mathcal{G}_{uv})|x_{uv}) =$$
$$= \frac{1}{|\mathcal{D}_{x_{uv}}|} \sum_{\{u_i, v_i\} \in \mathcal{D}_{x_{uv}}} K_{\Delta\lambda_2}(\{u,v\}, \{u_i, v_i\}) \quad (3)$$

where

$$K_{\Delta\lambda_2}(\{u,v\}, \{u_i, v_i\}) =$$
$$= \frac{1}{h} K\left(\frac{\Delta\lambda_2(u,v,\mathcal{G}_{uv}) - \Delta\lambda_2(u_i, v_i, \mathcal{G}_{u_i v_i})}{h}\right) \quad (4)$$

and $K$ is a kernel function with bandwidth $h$ [Silverman, 1986]. In our implementation of FRGs, we use the Epanechnikov kernel [Epanechnikov, 1969], namely $K(t) = \frac{3}{4}(1 - t^2)\mathbb{1}_1(|t|)$, where $\mathbb{1}_y(x) = 1$ if $x \leq y$, and $\mathbb{1}_y(x) = 0$ otherwise. The Epanechnikov kernel is more appealing than other possible functions (such as the Gaussian kernel) because of its computational convenience. At the same time, it displays analogous properties in terms of statistical consistency [Silverman, 1986].

## 3.3 DEPENDENCE STRUCTURE

In order to illustrate the behavior of FRGs with respect to conditional independence, we first recall the definition of Markov dependence for random graph models [Frank and Strauss, 1986]. Let $\mathcal{N}(X_{uv})$ denote the set $\{X_{wz}: \{w,z\} \in \mathcal{E} \land |\{w,z\} \cap \{u,v\}| = 1\}$. A Markov random graph is then defined as follows:

**Definition 2.** *A random graph $\mathcal{G}$ is said to be a* Markov graph *(or to have a* Markov dependence structure*) if, for any pair of variables $X_{uv}$ and $X_{wz}$ in $\mathcal{G}$ such that $\{u,v\} \cap \{w,z\} = \emptyset$, we have that $P(X_{uv}|X_{wz}, \mathcal{N}(X_{uv})) = P(X_{uv}|\mathcal{N}(X_{uv}))$.*

Based on Definition 2, we say that the dependence structure of a random graph $\mathcal{G}$ is *non-Markovian* if, for disjoint pairs of nodes $\{u,v\}$ and $\{w,z\}$, it does not imply that $P(X_{uv}|X_{wz}, \mathcal{N}(X_{uv})) = P(X_{uv}|\mathcal{N}(X_{uv}))$, i.e. if it is consistent with the inequality $P(X_{uv}|X_{wz}, \mathcal{N}(X_{uv})) \neq P(X_{uv}|\mathcal{N}(X_{uv}))$. We can then prove the following proposition:

**Proposition 3.** *There exist Fiedler random graphs with non-Markovian dependence structure.*

*Proof.* To prove the proposition, we make use of the following equalities [Fiedler, 1973]: if graphs $\mathcal{G}_1$ and $\mathcal{G}_2$ are, respectively, a path and a circuit of size $n$, then $\lambda_2(\mathcal{G}_1) = 2(1 - \cos(\pi/n))$ and $\lambda_2(\mathcal{G}_2) = 2(1 - \cos(2\pi/n))$. Consider a graph $\mathcal{G} = (\mathcal{V}, \mathcal{E})$ such that $\mathcal{V} = \{u, v, w, z\}$ and $\mathcal{E} = \{\{u,v\}, \{v,w\}, \{w,z\}, \{u,z\}\}$, and let $\mathcal{G}_{uv} = (\mathcal{V}_{uv}, \mathcal{E}_{uv})$ be the subgraph of $\mathcal{G}$ incident to $\mathcal{E}_{uv} = \mathcal{E} \setminus \{\{w,z\}\}$. If $\mathcal{G}$ is a Markov graph, it must be the case that $P(X_{uv}|X_{wz}, X_{vw}, X_{uz}) = P(X_{uv}|X_{vw}, X_{uz})$. Now, suppose that $\mathcal{G}$ is a FRG. In this case, we have that

$$P(X_{uv}|X_{wz}, X_{vw}, X_{uz}) = P(X_{uv}|\mathcal{X}(\mathcal{G}) \setminus \{X_{uv}\})$$
$$= \frac{p(\Delta\lambda_2(u,v,\mathcal{G})|X_{uv})P(X_{uv})}{p(\Delta\lambda_2(u,v,\mathcal{G}))} \quad (5)$$

and, similarly,

$$P(X_{uv}|X_{vw}, X_{uz}) = \frac{p(\Delta\lambda_2(u,v,\mathcal{G}_{uv})|X_{uv})P(X_{uv})}{p(\Delta\lambda_2(u,v,\mathcal{G}_{uv}))} \quad (6)$$

Eqs. 5–6 imply that, if the dependence structure of $\mathcal{G}$ is Markovian, then the following equality must hold:

$$\frac{p(\Delta\lambda_2(u,v,\mathcal{G})|X_{uv})}{p(\Delta\lambda_2(u,v,\mathcal{G}))} = \frac{p(\Delta\lambda_2(u,v,\mathcal{G}_{uv})|X_{uv})}{p(\Delta\lambda_2(u,v,\mathcal{G}_{uv}))} \quad (7)$$

Since the configuration of $\mathcal{G}$ and $\mathcal{G}_{uv}$ is given by a circuit and a path respectively, where both have size 4, we know that $\lambda_2(\mathcal{G}) = 2\,(1-\cos(\pi/2))$ and $\lambda_2(\mathcal{G}_{uv}) = 2\,(1-\cos(\pi/4))$. Also, if $\mathcal{G}^- = (\mathcal{V}, \mathcal{E} \setminus \{\{u,v\}\})$ and $\mathcal{G}_{uv}^- = (\mathcal{V}_{uv}, \mathcal{E}_{uv} \setminus \{\{u,v\}\})$, notice that $\lambda_2(\mathcal{G}^-) = \lambda_2(\mathcal{G}_{uv})$, since $\mathcal{G}^-$ is also a path of size 4, and that $M(0, \mathcal{G}_{uv}^-) = M(0, \mathcal{G}_{uv}) + 1$, since $\mathcal{G}_{uv}^-$ has one more connected component than $\mathcal{G}_{uv}$. Therefore, we have that $\Delta\lambda_2(u,v,\mathcal{G}) = 2\cos(\pi/4)$ and $\Delta\lambda_2(u,v,\mathcal{G}_{uv}) = 2\,(1-\cos(\pi/4))$, i.e. $\Delta\lambda_2(u,v,\mathcal{G}) \neq \Delta\lambda_2(u,v,\mathcal{G}_{uv})$. Because of this inequality, there will exist parameterizations of $p$ which do not satisfy Eq. 7, which means that the dependence structure of $\mathcal{G}$ is non-Markovian. □

Note that the proof of Proposition 3 can be straightforwardly generalized to the dependence between $X_{uv}$ and $X_{wz}$ in circuits/paths of arbitrary size $n$, since the expression used for the Fiedler eigenvalues of such graphs holds for any $n$. In fact, suppose that the 4-nodes circuit $\mathcal{G}$ used in the proof is replaced with a circuit $\mathcal{G}^* = (\mathcal{V}^*, \mathcal{E}^*)$ of size $n$, where $\mathcal{V}^* = \mathcal{V} \cup \{s_1, \ldots, s_1, \ldots, s_m, t_1, \ldots, t_m\}$ and $\mathcal{E}^*$ is obtained from $\mathcal{E}$ by replacing $\{u,z\}$ and $\{v,w\}$, respectively, with a path from $u$ to $z$ going through $s_1, \ldots, s_m$ and a path from $v$ to $w$ going through $t_1, \ldots, t_m$, so that $n = 2m + 4$. In this case, if $\mathcal{G}_{uv}^*$ is the subgraph of $\mathcal{G}^*$ incident to $\mathcal{E}_{uv}^* = \mathcal{E}^* \setminus \{\{w, z\}\}$, we have again that $\Delta\lambda_2(u,v,\mathcal{G}^*) \neq \Delta\lambda_2(u,v,\mathcal{G}_{uv}^*)$, which means that there exist FRGs such that $\frac{p(\Delta\lambda_2(u,v,\mathcal{G}^*)|\,X_{uv})}{p(\Delta\lambda_2(u,v,\mathcal{G}^*))} \neq \frac{p(\Delta\lambda_2(u,v,\mathcal{G}_{uv}^*)|\,X_{uv})}{p(\Delta\lambda_2(u,v,\mathcal{G}_{uv}^*))}$.

## 4 OTHER CONDITIONAL RANDOM NETWORKS

In this section, we adapt three families of random graph models—namely the ERG, Watts-Strogatz (WS), and Barabási-Albert (BA) models respectively—to the task of conditional estimation of edge distributions, and we also review the entailed dependence structures. This will allow for a clear comparison of such models to FRGs in Sec. 5. Kronecker graphs are not considered here because it is not straightforward how they could be turned into conditional estimators.

### 4.1 EXPONENTIAL RANDOM GRAPH MODELS

Let $G(\mathcal{V})$ be the set of all possible (undirected) graphs $\mathcal{G}$ with fixed vertex set $\mathcal{V}$, where $|G(\mathcal{V})| = \sqrt{2^{|\mathcal{V}|^2-|\mathcal{V}|}}$. Suppose that $\Phi$ denotes a collection of potential functions $\varphi_i$ over graphs, where each $\varphi_i(\mathcal{G})$ is a graph statistic such as the number of edges or the number of triangles in $\mathcal{G}$. Then, an ERG with parameter vector $\boldsymbol{\theta} = (\theta_1, \ldots, \theta_{|\Phi|})$ defines the following Boltzmann distribution [Robins et al., 2007]:

$$P(\mathcal{X}(\mathcal{G})|\,\boldsymbol{\theta}) = \frac{1}{Z(\boldsymbol{\theta})} \exp\left\{\sum_{i=1}^{|\Phi|} \theta_i \varphi_i(\mathcal{G})\right\} \quad (8)$$

where each $\theta_i$ is a real-valued parameter associated to the potential function $\varphi_i$, and $Z(\boldsymbol{\theta})$ denotes the partition function [Koller and Friedman, 2009], given by $Z(\boldsymbol{\theta}) = \sum_{\mathcal{G}_j \in G(\mathcal{V})} \exp\left\{\sum_{i=1}^{|\Phi|} \theta_i \varphi_i(\mathcal{G}_j)\right\}$. Computing the partition function exactly is clearly intractable for graphs with more than a few nodes, which makes it quite expensive to compute joint probabilities in large networks using ERGs.

ERGs entail a simple form for the conditional distribution of edges given the remainder of the graph:

$$P(X_{uv}|\,\mathcal{X}(\mathcal{G}) \setminus \{X_{uv}\}; \boldsymbol{\theta}) = \frac{\exp\left\{\sum\limits_{i=1}^{|\Phi|} \theta_i \varphi_i(\mathcal{G})\right\}}{\sum\limits_{x_{uv}} \exp\left\{\sum\limits_{i=1}^{|\Phi|} \theta_i \varphi_i(\mathcal{G}_{x_{uv}})\right\}} \quad (9)$$

where $\mathcal{G}_{x_{uv}}$ denotes the configuration of $\mathcal{G}$ that we obtain by clamping the value of $X_{uv}$ to $x_{uv}$. Therefore, if we want to estimate the parameters of an ERG so as to maximize the model log-likelihood with respect to a data sample $\mathcal{D} = \{(x_{u_1v_1}, \mathcal{G}_{\mathcal{S}_1}), \ldots, (x_{u_nv_n}, \mathcal{G}_{\mathcal{S}_n})\}$, we can optimize each parameter $\theta_i$ by taking the derivative $\frac{\partial}{\partial \theta_i} \sum_{j=1}^n \log P(x_{u_jv_j}|\,\mathcal{X}(\mathcal{G}_{\mathcal{S}_j}) \setminus \{X_{uv}\}; \boldsymbol{\theta})$ and applying any gradient-based optimization technique.

#### 4.1.1 Markov Random Graphs

Given a graph $\mathcal{G} = (\mathcal{V}, \mathcal{E})$, define a $k$-star (for $k \geq 2$) as a set $\mathcal{E}_{\mathcal{S}_k} \subseteq \mathcal{E}$ such that $|\mathcal{E}_{\mathcal{S}_k}| = k$ and, for any $\{u,v\}$ and $\{w,z\}$ in $\mathcal{E}_{\mathcal{S}_k}$, $\{u,v\} \cap \{w,z\} = 1$. Also, let a $k$-triangle (for $k \geq 1$) be a set $\mathcal{E}_{\mathcal{T}_k} \subseteq \mathcal{E}$ of size $2k+1$ such that the elements of $\mathcal{E}_{\mathcal{T}_k}$ are edges $\{u,v\}, \{u,w_1\}, \{v,w_1\}, \ldots, \{u,w_k\}, \{v,w_k\}$, where $w_i \neq w_j$ for any $i$ and $j$ such that $1 \leq i, j \leq k$. If we use $E(\mathcal{G})$, $S_k(\mathcal{G})$, and $T_k(\mathcal{G})$ to denote, respectively, the number of edges, the number of $k$-stars, and the number of $k$-triangles in $\mathcal{G}$, then an *exponential Markov random graph* (MRG) model with parameter vector $\boldsymbol{\theta}$ is defined by the following probability distribution [Frank and Strauss, 1986]:

$$P(\mathcal{G}|\,\boldsymbol{\theta}) = \frac{1}{Z} \exp\left\{\eta\, E(\mathcal{G}) + \sum_{k=2}^{K} \sigma_k S_k(\mathcal{G}) + \tau\, T(\mathcal{G})\right\} \quad (10)$$

where $K$ is the maximum value that we want to consider for $k$ in the $k$-star statistics, $\boldsymbol{\theta} = (\eta, \sigma_2, \ldots, \sigma_K, \tau)$, and $T(\mathcal{G}) = T_1(\mathcal{G})$ is simply the number of triangles in $\mathcal{G}$. That is to say, a MRG defines a Boltzmann distribution over graphs, with parameters corresponding to edge, $k$-star, and triangle statistics. Interestingly, the ER model can be characterized as a special case of MRG, where all the parameters except for $\eta$ are set to zero.

MRGs are known to satisfy Definition 2 [Robins et al., 2007]. Therefore, under the probability model defined by Eq. 10, we have that, for any pair of nodes $\{u,v\}$ and any subgraph sample $\mathcal{G}_\mathcal{S}$ from $\mathcal{G}$, $P(X_{uv}|\mathcal{X}(\mathcal{G}_\mathcal{S}) \setminus \{X_{uv}\}; \boldsymbol{\theta}) = P(X_{uv}|\mathcal{N}_\mathcal{S}(X_{uv}); \boldsymbol{\theta})$, where $\mathcal{N}_\mathcal{S}(X_{uv})$ is the set of all variables $X_{wz}$ from $\mathcal{G}_\mathcal{S}$ such that $|\{w,z\} \cap \{u,v\}| = 1$. Note that while Definition 2 specifies a general class of random graph models, MRGs in the strict sense refer to the class of ERGs defined above.

### 4.1.2 Higher-Order Models

Higher-order ERG models (HRGs) are readily obtained from MRGs by adding $k$-triangle counts (for $k \geq 1$) to the Boltzmann distribution of Eq. 10 [Robins et al., 2007]. Moreover, in order to avoid fixing in advance the maximum value of the $k$ parameter, a general formulation has been proposed for HRGs through the *alternating $k$-star* and the *alternating $k$-triangle* statistics [Snijders et al., 2006], obtaining the following probability model:

$$P(\mathcal{G}|\boldsymbol{\theta}) = \frac{1}{Z} \exp\left\{\eta E(\mathcal{G}) + \sigma S^*(\mathcal{G}) + \tau T^*(\mathcal{G})\right\} \quad (11)$$

where, if $n$ is the number of nodes in $\mathcal{G}$ and $S_k$ and $T_k$ are the usual $k$-stars and $k$-triangle counts, $S^*(\mathcal{G})$ and $T^*(\mathcal{G})$ are defined as $S^*(\mathcal{G}) = \sum_{k=2}^{n-1} (-1)^k \frac{S_k}{\rho^{k-2}}$ and $T^*(\mathcal{G}) = \sum_{k=2}^{n-2} (-1)^k \frac{T_k}{\rho^{k-2}}$ (with $\rho \geq 1$ acting as a sort of regularization parameter).

It is known that, under the distribution given by Eq. 11, $P(X_{uv}|\mathcal{X}(\mathcal{G}_\mathcal{S}) \setminus \{X_{uv}\}; \boldsymbol{\theta}) = P(X_{uv}|\mathcal{N}_\mathcal{S}^*(X_{uv}); \boldsymbol{\theta})$ [Robins et al., 2007], where $\mathcal{N}_\mathcal{S}^*(X_{uv})$ is the set containing any $X_{wz}$ from $\mathcal{G}$ such that $X_{wz} \neq X_{uv}$ and, for at least one edge $\{s,t\}$, we have that $|\{s,t\} \cap \{u,v\}| = 1$ and $\{s,t\} \cap \{w,z\} \neq \emptyset$. Clearly, $\mathcal{N}_\mathcal{S}(X_{uv}) \subseteq \mathcal{N}_\mathcal{S}^*(X_{uv})$, which is why HRGs are 'higher-order' than MRGs.

## 4.2 WATTS-STROGATZ MODEL

The WS network model [Watts and Strogatz, 1998] defines a random network as a regular ring lattice which is randomly 'rewired' so as to introduce a certain amount of disorder, which typically leads to small-world phenomena. Given nodes $u_1, \ldots, u_n$, a WS network is generated by constructing a regular ring lattice such that each node is connected to exactly $2\delta$ other nodes. Network edges are then scanned sequentially, and each one of them is rewired with probability $\beta$, where, if $i < j$, rewiring an edge $\{u_i, u_j\}$ means to replace it with another edge $\{u_i, u_k\}$ such that $k \neq i$ and $u_k$ is chosen uniformly at random from the set of all nodes that are not already linked to $u_i$.

Interestingly, the degree distribution corresponding to a WS network $\mathcal{G}$ with parameters $\delta$ and $\beta$ takes the following form [Barrat and Weigt, 2000], for any degree $k \geq \delta$:

$$P(k) = \sum_{i=0}^{I} \binom{\delta}{i} (1-\beta)^i \beta^{\delta-i} \frac{(\delta\beta)^{k-\delta-i}}{(k-\delta-i)!} \exp(-\beta\delta) \quad (12)$$

where $I = \min\{k-\delta, \delta\}$. Given Eq. 12, we model the conditional distribution of a variable $X_{uv}$ given the remainder of $\mathcal{G}$ through the following quantity:

$$P(X_{uv}|\mathcal{X}(\mathcal{G}) \setminus \{X_{uv}\}) = \frac{P(d_u(\mathcal{G})) \, P(d_v(\mathcal{G}))}{\sum_{x_{uv}} P(d_u(\mathcal{G}_{x_{uv}})) \, P(d_v(\mathcal{G}_{x_{uv}}))} \quad (13)$$

where $d_u(\mathcal{G})$ denotes the degree of node $u$ in $\mathcal{G}$, and each $P$ is implicitly conditional on the values of $\delta$ and $\beta$. We refer to the conditional random graph model specified in Eq. 13 as the *conditional Watts-Strogatz (CWS) model*.

For the CWS model, the following proposition follows straightforwardly from Eq. 13:

**Proposition 4.** *The dependence structure of CWS models is Markovian.*

A simple estimate of the $\delta$ parameter for a data sample $\mathcal{D} = \{(x_{u_1 v_1}, \mathcal{G}_{u_1 v_1}), \ldots, (x_{u_n v_n}, \mathcal{G}_{u_n v_n})\}$ is the following:

$$\hat{\delta} = \frac{1}{2n} \sum_{i=1}^{n} d_{u_i}(\mathcal{G}_{\mathcal{S}_i}) + d_{v_i}(\mathcal{G}_{u_i v_i}) \quad (14)$$

On the other hand, a simple strategy for estimating the rewiring probability by a maximum likelihood approach consists in parameterizing it as a sigmoid function $\beta = \frac{1}{1+\exp(-\theta_\beta)}$, where $\theta_\beta$ can be optimized by exploiting the derivative $\frac{\partial}{\partial \theta_\beta} \sum_{i=1}^{n} \log P(x_{u_i v_i}|\mathcal{X}(\mathcal{G}_{u_i v_i}) \setminus \{X_{uv}\}; \delta, \theta_\beta)$.

## 4.3 BARABÁSI-ALBERT MODEL

The BA model was originally proposed for explaining the scale-free degree distributions often observed in real-world networks [Barabási and Albert, 1999]. In the BA model, the probability $P(u)$ of linking to any particular node $u$ in a network $\mathcal{G} = (\mathcal{V}, \mathcal{E})$ takes the form $P(u) = \frac{d_u(\mathcal{G})^\alpha}{\sum_{v \in \mathcal{V}} d_v(\mathcal{G})^\alpha}$ [Albert and Barabási, 2002], where $\alpha$ is a real-valued parameter affecting the shape of the degree distribution. Given $P(u)$, we can use the following expression to characterize the conditional probability of observing edge $\{u,v\}$ given the

remainder of $\mathcal{G}$ [Newman, 2001, Barabási et al., 2002]:

$$P(X_{uv} = 1 | \mathcal{X}(\mathcal{G}) \setminus \{X_{uv}\}; \alpha) = \frac{d_u(\mathcal{G})^\alpha \, d_v(\mathcal{G})^\alpha}{\left(\sum_{w \in \mathcal{V}} d_w(\mathcal{G})^\alpha\right)^2} \tag{15}$$

We refer to the random graph model of Eq. 15 as the *conditional Barabási-Albert (CBA) model*. The $\alpha$ parameter can be optimized straightforwardly by gradient-based methods, so as to maximize the objective $\sum_{i=1}^n \log P(x_{u_i v_i} | \mathcal{X}(\mathcal{G}_{u_i v_i}) \setminus \{X_{uv}\}; \alpha)$ over an observed sample of size $n$.

The following property holds for CBA networks:

**Proposition 5.** *The dependence structure of CBA models is non-Markovian.*

*Proof.* Consider a CBA network $\mathcal{G} = (\mathcal{V}, \mathcal{E})$, where $\{u, v\}$ and $\{w, z\}$ are edges in $\mathcal{E}$ such that $\{u, v\} \cap \{w, z\} = \emptyset$. $\mathcal{G}$ is a Markov graph if and only if $P(X_{uv} | X_{wz}, \mathcal{N}(X_{uv})) = P(X_{uv} | \mathcal{N}(X_{uv}))$. Let $\mathcal{G}_1$ and $\mathcal{G}_2$ be the same as defined in the proof of Proposition 4. Although $d_u(\mathcal{G}_2) = d_u(\mathcal{G}_1)$ and $d_v(\mathcal{G}_2) = d_v(\mathcal{G}_1)$, we have that $\sum_{w \in \mathcal{V}} d_w(\mathcal{G}_2)^\alpha > \sum_{w \in \mathcal{V}} d_w(\mathcal{G}_1)^\alpha$. Therefore,

$$\begin{aligned} P(X_{uv} | X_{wz}, \mathcal{N}(X_{uv})) &= P(X_{uv} | \mathcal{X}(\mathcal{G}_2) \setminus \{X_{uv}\}) \\ &= \frac{d_u(\mathcal{G}_2)^\alpha \, d_v(\mathcal{G}_2)^\alpha}{\left(\sum_{w \in \mathcal{V}} d_w(\mathcal{G}_2)^\alpha\right)^2} \neq \frac{d_u(\mathcal{G}_1)^\alpha \, d_v(\mathcal{G}_1)^\alpha}{\left(\sum_{w \in \mathcal{V}} d_w(\mathcal{G}_1)^\alpha\right)^2} \\ &= P(X_{uv} | \mathcal{X}(\mathcal{G}_1) \setminus \{X_{uv}\}) = P(X_{uv} | \mathcal{N}(X_{uv})) \end{aligned} \tag{16}$$

□

## 5 EXPERIMENTAL EVALUATION

In order to assess experimentally the accuracy of FRGs as models of conditional edge distributions in large-scale networks, we test them in the following link prediction setting. First, we take a network $\mathcal{G} = (\mathcal{V}, \mathcal{E})$, with a number of nodes ranging from a few thousands to a couple of millions, and a number of edges going up to several millions. We then sample $n$ training pairs $\{u_i, v_i\}$ uniformly at random from $\mathcal{V}$, with $u_i \neq v_i$. For each sampled pair of nodes $\{u_i, v_i\}$, we recover the subgraph $\mathcal{G}_{u_i v_i}$ induced on $\mathcal{G}$ by the vertex set $\mathcal{V}_{u_i v_i} = \{u_i, v_i\} \cup \{w_i : \{u_i, w_i\} \in \mathcal{E} \lor \{v_i, w_i\} \in \mathcal{E}\}$, so as to get a training set $\mathcal{D} = \{(x_{u_1 v_1}, \mathcal{G}_{u_1 v_1}), \ldots, (x_{u_n v_n}, \mathcal{G}_{u_n v_n})\}$. We then estimate from $\mathcal{D}$ each one of the conditional random graph models described in the previous sections, using the respective learning techniques. Given the estimated models, we sample a test set $\mathcal{T} = \{(x_{u_1 v_1}, \mathcal{G}_{u_1 v_1}), \ldots, (x_{u_m v_m}, \mathcal{G}_{u_m v_m})\}$ from $\mathcal{G}$ such that $\mathcal{T} \cap \mathcal{D} = \emptyset$, where each element of $\mathcal{T}$ is sampled through the same technique used for $\mathcal{D}$. For each estimated model, we assess its accuracy by first using it to compute the conditional probability of observing a link for each pair of nodes $\{u_j, v_j\}$ in $\mathcal{T}$, and then calculating the area under the ROC curve (AUC) for the predictions delivered on $\mathcal{T}$. Since the problem of predicting the presence of edges in real-world networks is severely unbalanced, i.e. $|\mathcal{E}| \ll \frac{1}{2}(|\mathcal{V}|^2 - |\mathcal{V}|)$, the ROC curve is a particularly appropriate evaluation measure [Fawcett, 2006]. Also, in order to assess the sensitivity of each model to the size of the training sample, we ran some preliminary experiments with several different values of $|\mathcal{D}|$. The result of this preliminary investigation was that, provided that the training sample contains at least a few pairs of linked nodes (in the order of ten or twenty), then the behavior of the different models is not significantly affected by increasing sample sizes.

Table 1: General statistics for the network datasets used in the experiments. $CC_\mathcal{G}$ and $D_\mathcal{G}$ denote the average clustering coefficient and the network diameter respectively.

| **Network** | $|\mathcal{V}|$ | $|\mathcal{E}|$ | $CC_\mathcal{G}$ | $D_\mathcal{G}$ |
|---|---|---|---|---|
| **AstroPh** | 18,772 | 396,160 | 0.63 | 14 |
| **GrQc** | 5,242 | 28,980 | 0.52 | 17 |
| **HepPh** | 12,008 | 237,010 | 0.61 | 13 |
| **HepTh** | 9,877 | 51,971 | 0.47 | 17 |
| **Enron** | 36,692 | 367,662 | 0.49 | 12 |
| **RoadCA** | 1,965,206 | 5,533,214 | 0.04 | 850 |
| **RoadTX** | 1,379,917 | 3,843,320 | 0.04 | 1,049 |

We compare the performance of the MRG, HRG, CWS, CBA, and FRG models on seven different networks, drawn from the arXiv e-print repository, the Enron email corpus, and the US road network respectively [Leskovec et al., 2007, 2009]. The ER model is not considered in these experiments, since the involved independence assumption makes that model simply unusable for conditional estimation tasks. Some representative network statistics for the used datasets are summarized in Table 1. For the arXiv datasets (i.e. the AstroPh, GrQc, HepPh, and HepTh networks), we report results for a choice of 10,000 node pairs both for the training and the test set, whereas for the Enron and US road (RoadCA and RoadTX) networks we have $|\mathcal{D}| = |\mathcal{T}| = 100,000$ and $|\mathcal{D}| = |\mathcal{T}| = 10,000,000$ respectively. However, notice that nearly identical results can be obtained in each case after reducing the number of training samples down to even 20%–40% of the indicated size (as discovered through preliminary analysis). ROC curves for the described datasets are plotted in Figs. 1–3, whereas the corresponding AUC values are given in Table 2. As a general remark, we point out that AUC values less than 0.5 (as recorded for some models) do not necessarily mean that prediction accuracy is being worse than random guessing. In fact, recent studies show that, depending on signal strength and stratification errors between training and test sets, random guessing can drop to AUC values even lower than 0.3 [Parker et al., 2007]. Therefore, AUC values less than 0.5 should be considered simply as *no better* than random guessing, rather than *worse* than random.

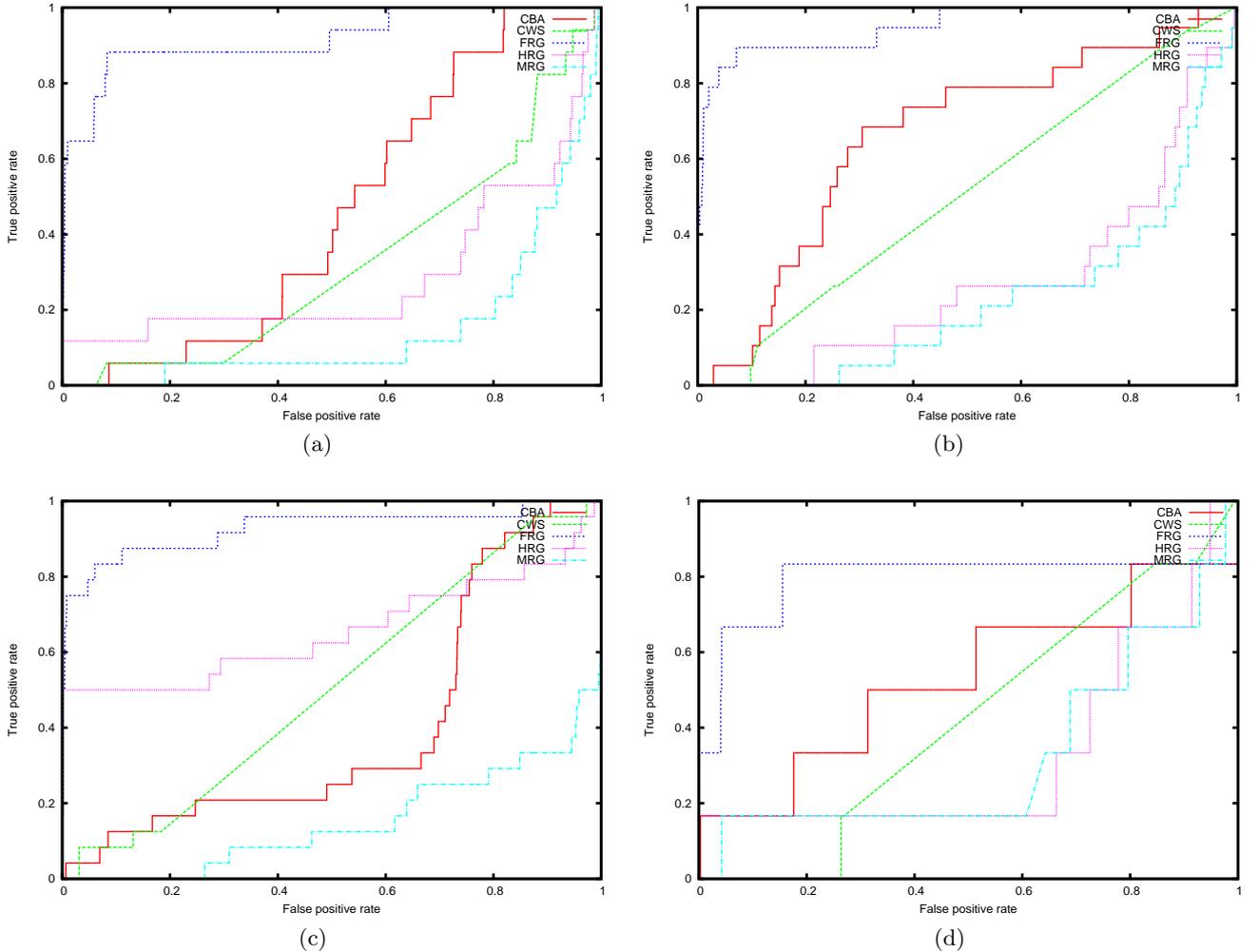

Figure 1: ROC curves for the AstroPh (a), GrQc (b), HepPh (c), and HepTh (d) networks.

FRGs significantly outperform all other models both in the arXiv and in the Enron networks. Such networks are generally characterized by power law degree distributions, contrary to the road networks, which exhibit instead a regular topology, with degree distribution peaked around its mean. This explains the better behavior of the CBA model with respect to the remaining options in most of the scale-free networks. Interestingly, this trend is reverted for the HepPh data, where not only CWS outperforms CBA, but we also observe that HRG behaves much better than it does on the other scale-free networks. The fact that, in this particular case, FRGs remain the best available option provides evidence for their higher stability (i.e. robustness to variation in the underlying network properties) with respect to the other models. On the other hand, for the US road networks we observe a completely different pattern of results. This time, the FRG model performs worse than the CWS model or the two variants of ERG. The reason is related to the lattice-shaped regularity exhibited by such graphs. This is not surprising, as the WS model subsumes regular lattices as special case scenarios. A natural explanation for the superiority of the CWS model on the road networks lies in the fact that their parametric assumption concerning the observed degree distribution happens to be satisfied for such data, while it is clearly violated by the scale-free networks, which is exactly the opposite of what happens for the CBA model. It is worth noting, however, that FRGs are again performing significantly better than the CBA model. Therefore, FRGs appear to be a more stable class of conditional random network models, since their predictions ensure a regularly accurate behavior across networks displaying significantly different statistical properties. Overall, the presented results encourage the hypothesis that a genuinely nonparametric approach to conditional random graph modeling can offer dramatic advantages over parametric approaches.

Table 2: AUC values for the ROC curves plotted in Figs. 1–3. $CC_\mathcal{G}$ and $D_\mathcal{G}$ denote the average clustering coefficient and the network diameter respectively.

| | AUC (%) | | | | |
|---|---|---|---|---|---|
| **Network** | **CBA** | **CWS** | **FRG** | **HRG** | **MRG** |
| **AstroPh** | 46.04 | 32.39 | **91.66** | 28.69 | 14.78 |
| **GrQc** | 66.22 | 50.86 | **94.90** | 27.10 | 22.32 |
| **HepPh** | 40.04 | 50.84 | **92.73** | 65.62 | 15.02 |
| **HepTh** | 53.21 | 42.47 | **79.34** | 32.82 | 32.36 |
| **Enron** | 48.45 | 49.03 | **85.92** | 19.94 | 11.23 |
| **RoadCA** | 30.84 | **91.21** | 66.70 | 77.90 | 81.12 |
| **RoadTX** | 34.67 | **93.01** | 63.75 | 62.51 | 68.36 |

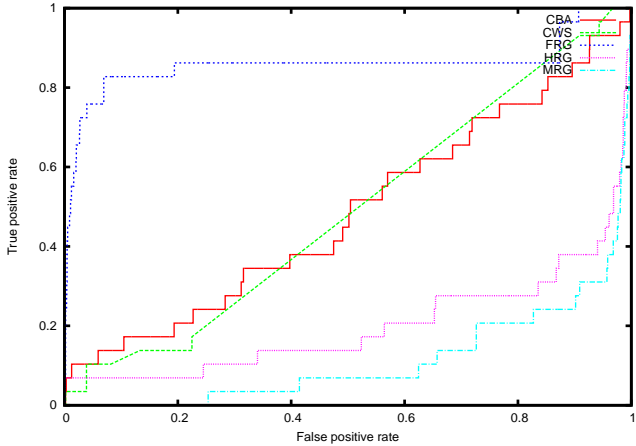

Figure 2: ROC curves for the Enron network.

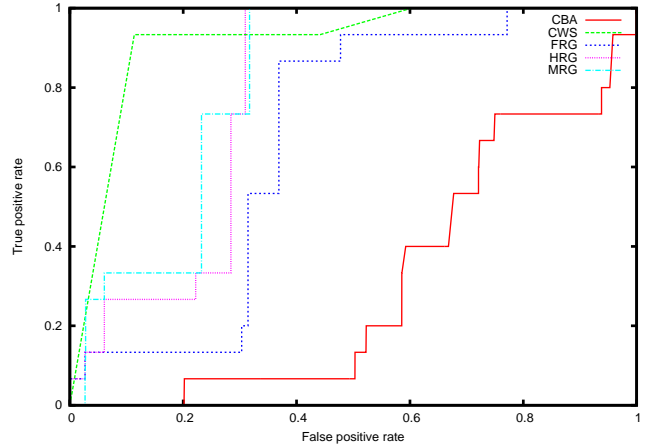

(a)

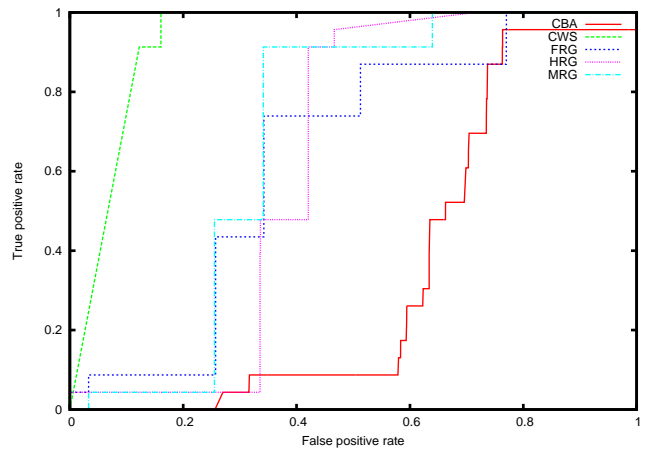

(b)

Figure 3: ROC curves for the RoadCA (a) and RoadTX (b) networks.

## 6 CONCLUSION

The work presented in this paper started from two motivations. On the one hand, we remarked that statistical modeling of networks cries for nonparametric estimation, because of the inaccuracy often resulting from fallacious parametric assumptions. In this respect, we showed that statistics derived from the Laplacian spectrum of graphs (such as the Fiedler delta function) offer practical ways of developing nonparametric estimators. On the other hand, we suggested that a conditional approach to random graph modeling can be very effective in the large-scale setting, since it allows e.g. to dispense with the intractable partition functions often involved in joint distributions. With respect to this point, we showed how FRGs allow us to tackle very large datasets, processing effectively even millions of queries over networks with millions of nodes and edges. Interesting options for future work consist in analyzing the effect of the subgraph sampling mechanism in determining the distribution of the Fiedler delta, as well as investigating alternative ways of exploiting the information enclosed in the graph spectrum.

**Acknowledgements**

This work has been supported by the French National Research Agency (ANR-09-EMER-007). The authors are grateful to Romaric Gaudel, Rémi Gilleron, and Michal Valko for their suggestions and comments concerning this work.